\documentclass[superscriptaddress, reprint, amsmath, amssymb, prx]{revtex4-2}
\usepackage{graphicx}
\usepackage[utf8]{inputenc}
\usepackage[dvipsnames]{xcolor}

\begin{document}

\title{Solid State Neuroscience: Spiking Neural Networks as Time Matter}
%\author{Jiaming Wu}
%\affiliation{Universit\'e Paris-Saclay, CNRS
%Laboratoire de Physique des Solides, 91405, Orsay, France}
\author{Marcelo Rozenberg}
\affiliation{Universit\'e Paris-Saclay, CNRS
Laboratoire de Physique des Solides, 91405, Orsay, France}

\date{\today}

\begin{abstract}
We aim at building a bridge between to {\it a priori} disconnected fields: Neuroscience
and Material Science. 
We construct an analogy based on identifying spikes events
in time with the positions of particles of matter.
We show that one may think of the dynamical states of spiking neurons and spiking neural
networks as {\it time-matter}. Namely, a structure of spike-events in time 
having analogue properties to that of ordinary matter.
%and material particles, where the time of spike events are associated to the position 
%of particles.
%In the simple case of a tonic spiking neuron, the inter-spike interval
%is associated to the lattice constant of a crystal, and the input excitation current
%is associated to the applied pressure.  
We can define for neural systems notions equivalent to the equations of state, 
phase diagrams and their phase transitions. For instance, the 
familiar Ideal Gas Law relation (P$v$ = constant)
emerges as analogue of the Ideal Integrate and Fire neuron model
relation ($I_{in}$ISI = constant). 
%frequency proportional to input current).
We define the neural analogue of the spatial structure correlation function,
that can characterize spiking states with temporal long-range order, such as regular tonic spiking.
%We extend the analogy to bursting neurons and show that the transition to
%a bursting state is analogous to proliferation and clustering of vacancy defects in a crystal.
%We also extend the analogy for neural networks, where each neuron 
%correspond to a different type of atom. Thus, recurrent 
%spike sequences play the role of an atomic basis that decorates 
%a Bravais lattice, or of a molecular motif in a molecular solid.
We also define the ``neuro-compressibility'' response function 
in analogy to the 
lattice compressibility. We show that similarly to the case of ordinary matter, 
the anomalous behavior of the neuro-compressibility is a precursor effect that
signals the onset of changes in spiking states.
We propose that the notion of neuro-compressibility may open the way to 
develop novel medical tools for the early diagnose of diseases. It may allow
to predict impending anomalous neural states, such as Parkinson's tremors, epileptic seizures, 
electric cardiopathies, and perhaps may even serve as a predictor of the likelihood
of regaining consciousness.
\end{abstract}
\maketitle

\section{Introduction}
The understanding of the mind is, arguably, the most mysterious scientific frontier. 
The ability of the mind to understand itself is puzzling. Nevertheless, it 
seems increasingly possible and within our reach.
Neuroscience and Artificial Intelligence are making great progress in that regard, however,
following very different paths and driven by very different motivations. 
In the first, the focus is to address fundamental questions of biology, 
while in the second, it is to develop brain inspired computational systems 
for practical applications for modern life. 
Evidently, there is also large overlap between them both.

The basic units that conform the physical support of the mind, namely the brain and the associated
neuronal systems, are neurons. These are cells with electrical activity that interact via electric
spikes, called action potentials \cite{neuroscience}. 
In animals, neurons conform networks
of a wide range of complexity. 
Ranging from a few hundred units in jelly fish, to a 
hundred billion in humans. 
A fundamental question to answer is {\it how and why} 
nature has adopted this electric signaling system. Its main functions
are multifold: to sense and monitor the environment, then to produce  
behavior and decision making, and finally to drive the required motor 
actions to assure the survival of living beings.
Neuroscience has already provided a good understanding of the 
electric behavior of individual neurons \cite{gerstner,ermentrout}. 
A major milestone was the explanation 
by Hodgkin and Huxley of the physiological mechanism for the generation of
the action potential \cite{hodgkin}. 
At the other end, that of neural networks with large number of neurons, 
important contributions come from Artificial Intelligence. 
For instance, significant progress was made in the 80s following the 
pioneering work of Hopfield \cite{hopfield1982neural}. 
More recently, this area received a renewed 
boost of activity, enabled by 
the combination of new learning algorithms for Deep Convolutional Neural Networks \cite{DeepNN} 
with the numerical power of modern computers \cite{merolla2014million}.
However, the networks adopted in Artificial Intelligence overwhelmingly 
describe the neurons' activity by their firing rate and not by individual spikes.
These are conceptually different, a spike is a discrete event, while the spiking rate 
is a continuous variable. Hence, modeling neurons in terms of the latter 
does not directly address the question posed above, namely, the why and how of 
Nature using discrete spikes.

Here we propose to look at this problem under a different light, 
which to my knowledge has not been discussed before. Hopefully, this may bring
new insights and perhaps help to develop our intuition for 
the challenging problem of understanding the mechanism of
spiking neural networks. We shall postulate 
an analogy between matter states, as in organized spatial patterns of particles
(or atoms or molecules) and that of neuronal states, as organized patterns of 
spikes in time.
Since we are attempting to build a bridge between
disconnected disciplines, we shall keep our presentation pedagogical.

Ultimately, as with any definition, our analogy will be of value 
if it turns to be a useful besides the intrinsic academic interest.
With this in mind, we shall later discuss an exciting perspective that 
the present approach may perhaps open. Namely, to develop novel screening 
tools that could allow to detect an enhanced risk to develop 
neural diseases that involve anomalous spiking
states, such as Parkinson's, epilepsy, cardiopathies, unconsciousness.

\section{The analogy}
As mentioned above, here we postulate an analogy between spatial matter states, 
such as solid, liquid and gas, and the dynamical states of spiking neurons and
neural networks. 
More specifically, an analogy between the organization of matter in space and
that of spiking events in time, which we call {\it temporal-matter} states.

\begin{figure}
%[htbp]
    \centering
    \includegraphics[width=\linewidth]{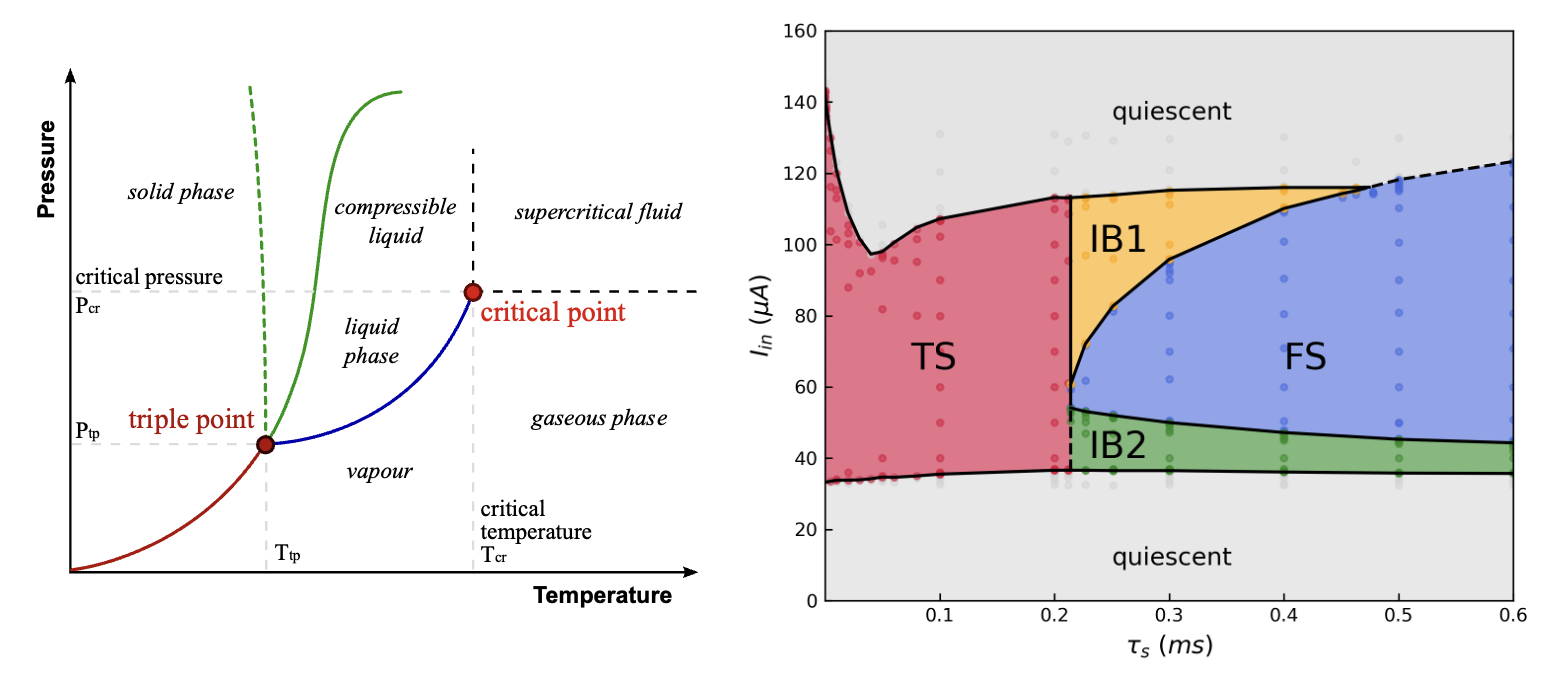}
    \caption{Left: Phase diagram of states of ordinary matter (water) showing the 
    solid, liquid and gas phases. $P$ and $T$ determine the specific volume $v$ of 
    the state,
    which is the inverse of the density $n$. Right: Measured phase diagram of 
    {\it temporal-matter} states of
    an electronic neuron model. The system is a two-compartment 
    bursting neuron, showing tonic spiking (TS), fast spiking (FS) and two different intrinsic 
    bursting states (IB1 and IB2). The input current $I$ and the circuit time-constant $\tau_s$
    determine the inter-spike interval of the state, which corresponds to the inverse of the frequency. 
    }
    \label{fig1}
\end{figure}

To motivate this, 
%We show that similarly to characterizing ordinary
%matter states by charting the ($p$,$v$) phase diagrams, we may explore and characterize
%neural states by charting the ($I$,$f$) phase diagram of a neural network system. 
%To make the analogy more precise, 
in Fig.\ref{fig1} we show the familiar phase diagram of
water as function of pressure and temperature ($p$,$T$). 
%Here the temperature acts as a 
%third parameter, which determines the specific volume. 
Next to it we show another phase diagram, that of an electronic 
bursting spiking neuron, which we introduce recently \cite{MBN}. 
In the diagram, we observe
various phases that correspond to qualitatively different states:
tonic spiking (TS), fast spiking (FS), and two types of bursting (IB1, IB2). 
The phase diagram is obtained as a function of two parameters, the excitatory 
current $I$
and a circuit time constant $\tau_s$ (the circuit is shown in Fig.\ref{fig4}).

%In the neural network temporal
%matter analogy, we may have applied current $I$ playing the role of pressure (more below) 
%and some model parameter such the time constant $\tau_s$ playing the role of $T$. 
%Thus, in the phase diagram ($I$,$\tau_s$) the coordinates define the temporal state of  
%system, i.e. its spiking state. 
To characterize the spiking states in the phase diagram, we need to consider the nature of the discrete
spiking events in the time domain. For instance, tonic spiking is characterized by a 
sequence of spikes that occur at equally spaced time-intervals. In Neuroscience, 
the time between two consecutive spike events is called the inter-spike interval (ISI). At the transition from the TS to FS phase, one observes a 
sudden decrease of the ISI (i.e. a jump in the spiking frequency) 
as function of the parameter $\tau_s$. 

 The ISI(t) characterizes the organization of the spikes in the time domain and
 it indicates the ``time-distance'' between spikes. If a sequence of spike
 events at times $t_i$ are indicated by the function 
 \begin{equation}
     s(t) = \sum_i \delta(t-t_i)
     \label{s}
 \end{equation}
 Then, it may be tempting to establish the analogy by simply identifying the
 time-position of spike events with the positions of particles in a matter state.
 The matter state is characterized by its particle density function,
 \begin{equation}
     n(x) = \sum_i \delta(x-x_i)
 \end{equation}
 where $x_i$ denotes the position of the $i^{th}$ particle. For simplicity,
 we assume point particles and one dimensional space. Similarly, in Eq.\ref{s}
 we have assume ideal spikes represented by $\delta$-functions, while in reality
 the action potentials have a typical duration of $\sim 1ms$
 From the analogy, the simple 
 tonic spiking case with equispaced spikes (i.e. constant ISI) 
 would correspond to a perfect crystal of equally spaced particles (atoms). 
 Thus, the familiar ISI of neuroscience would be the analogue 
 of the familiar lattice constant $a$ (or specific volume $v$) for material science.
 
 If one applies a pressure $P$ to matter, in general one observes
the decrease of $v$, such as in the case of the Ideal Gas law
 $P$$v$ = constant. 
 On the other hand, in neuroscience, it is well known that the ISI 
 can be reduced by increasing the excitatory input current. 
 Hence, one may be tempted to extend our analogy to associate the
 $P$ with the input $I$ in neural systems.
 We can make this more precise by introducing the simplest theoretical model 
 of an ideal spiking neuron, the Integrate and Fire (IF) model \cite{gerstner}. 
 
\begin{figure}
    \centering
    \includegraphics[width=0.8\linewidth]{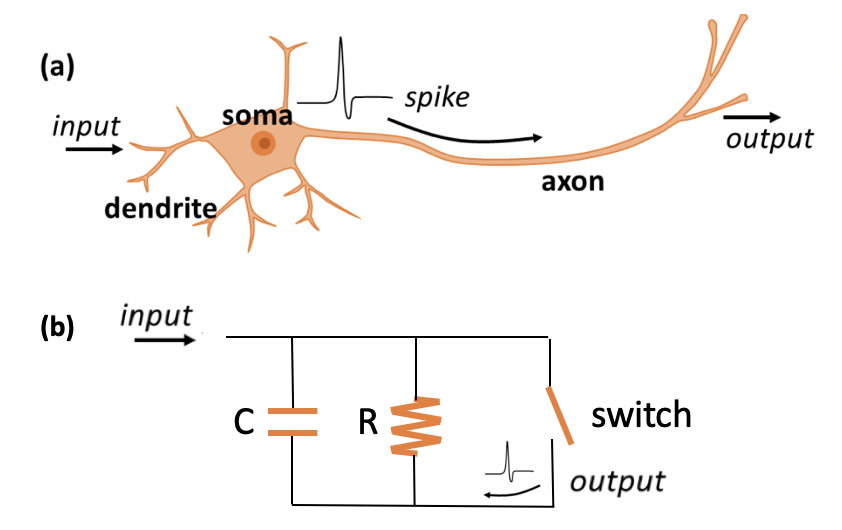}
    \caption{(a) Schematic biological neuron. 
    (b) Electric circuit of the IF model (for the case
    $u_{res}$=0). The membrane of the cell's body (soma) 
    is represented by the capacitor $C_m$, which accumulates
    charge of the input current. If the resistor $R \to \infty$ then the current integration
    is perfect, however for a finite value there is a ``leaky'' integration and the model is
    called LIF, for leaky-integrate-and-fire.}
\label{fig2}
\end{figure}

 In a very schematic view, as shown in Fig.\ref{fig2}, a biological neuron is 
composed of three main parts: the dendrites, the soma and the axon. The neuron 
is excited through the input of electric signals arriving to the dendrites, which 
are called the synaptic currents. This input is integrated in the 
cell's body, which leads to the increase in its electric potential with respect
to its resting state. By means of intense excitation, 
the neuron eventually reaches a threshold potential value. 
At that point a dramatic event takes place, an electric spike is initiated,
it propagates down the axon and eventually is 
communicated to the dendrites of a downstream
neuron. This phenomenon is called the emission of an action potential,
which was first described by Hodgkin and Huxley \cite{hodgkin}. 
This qualitative description can be
represented by the simple (leaky) IF model \cite{gerstner,ermentrout}
\begin{equation}
\frac{du}{dt} = -\frac{1}{\tau} u(t) + I; \ \ if \ u \geq u_{th} \ then \ spike \  and \ u = u_{rest}
\end{equation}
where $u(t)$ represents the potential of the soma, $u_{th}$ is the 
threshold potential,
$u_{rest}$ is the resting potential value and $I$ is the input (synaptic) 
current. The time constant $\tau$ is a characteristic
relaxation time of the neuron that represents the 
leakage of charge out
of the soma. If the leakage is negligible, $\tau \to \infty$, and the integration
is perfect, so one has an ideal IF model.
The electric circuit representation of the model is straight forward. 
The soma is represented
by a capacitor $C$ that accumulates the charge of the input current, and the 
leakage by a resistor $R$ in parallel. The threshold voltage can be represented 
by a switch that closes yielding the emission of the spike, which is the fast
discharge of the charge accumulated in $C$, as a delta function of current.

The limit of zero leakage, i.e. the ideal Integrate and Fire is trivially solved. 
The 
potential $u$ due to the integrated charge in $C$ during an interval $t$ is 
$u(t) = Q/C= (I/C)t$. Thus, the spike fire time $t_f$ is given by the condition 
$u(t_f)= u_{th}$. Which leads to $I t_f = u_{th} C$, or $I {\rm ISI} = constant$.

Hence, we may extend our analogy by noting that the equation of state
of the Ideal Gas has the same form as that of an ideal IF neuron, namely,

\begin{equation}
Pv {\rm = constant}   \longleftrightarrow I{\rm ISI = constant}
\label{eq:ideal}
\end{equation}

where 
\begin{equation}
P \longleftrightarrow I ;\ \ \ v \longleftrightarrow {\rm ISI}
\label{eq:equiv}
\end{equation}

As the equation of state of a real gas departs from the ideal case, 
biological neurons and neuron models' equation of state,  will depart
from the ideal IF ``neuronal equation of state'' above. 
Interestingly, the notion of neuronal equation of state should not appear 
so strange. Indeed, it is nothing else 
than the familiar concept in neuroscience of 
the neuron's {\it activation function}, namely,
the firing rate as a function of the excitatory input current, $f = f(I)$.
For instance, some popular models are: the rectified 
linear units, or ReLU, where $f(I) = max[0,I]$; the sigmoid activation 
$f(I) = 1/(1+e^{-I})$; etc.
These are examples of {\it mathematical} neuron models, however, we may also 
include here {\it physical} neuron models, namely, models that are defined
by an electronic circuit. In physical neuron models the equation of state
$f(I)$ can be measured, as in real gases. 
We should mention that while the relation $Pv$= constant is familiar from
the Ideal Gas law, it is in general valid for any liquid or solid in the
linear regime.

Recently, we introduced a {\it minimal} model of a physical
spiking neuron, which is achieved by exploiting the memristive properties of 
an ``old'' conventional electronic component, the thyristor. The model 
is minimal because it provides a physical realization of the basic
{\it leaky-integrate-and-fire} (LIF) neuron model by associating exactly one component
to each one of the three functions: a capacitor to integrate, a resistor to leak and 
the thyristor to fire. Qualitatively, the thyristor acts as the switch in the 
circuit of
Fig.\ref{fig2}.

\begin{figure}
    \centering
    \includegraphics[width=1\linewidth]{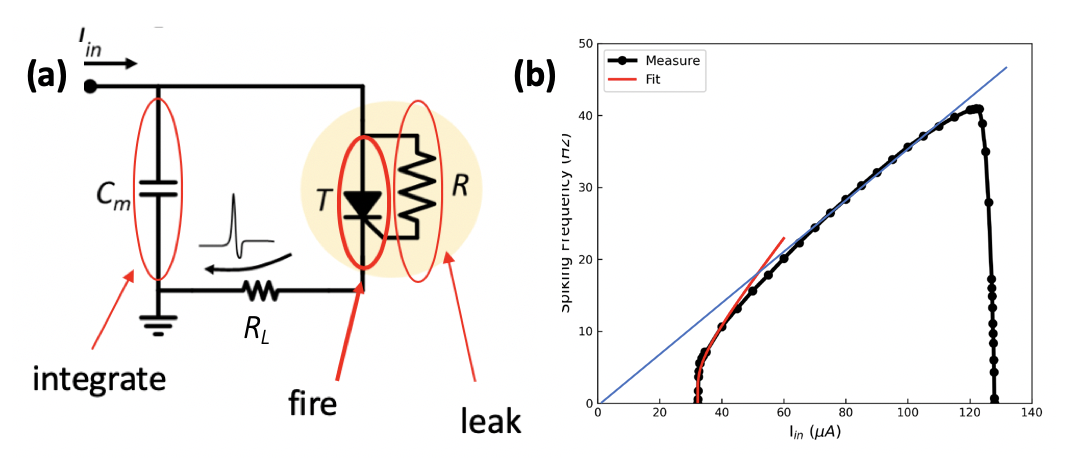}
    \caption{(a) Electric circuit that defines the physical neuron model based on the 
    memristive properties of the thyristor. The function of the thyristor is that 
    of a voltage controlled switch, as shown in Fig.\ref{fig2} above. The small 
    ``load'' resistor $R_L$ is to transform the output spike current into a voltage
    action potential signal for measurement convenience.
    (b) The ``neuron equation of state'' or
    activation function $f(I)$. Near the threshold, the equation of state follows closely
    the functional form of a LIF model (red line), ~$-1/[log(1 - I_c/I)]$ where
    $I_c$ is the minimal activation current. The blue line corresponds
    to the Ideal IF behavior $I \propto $ 1/ISI$\ =f$. 
    }
\label{fig3}
\end{figure}

In Fig.\ref{fig3} we show the circuit that defines physical neuron model just described, 
where we identify the role of each of the three components to the functions of the LIF. 
We call this artificial neuron model the Memristive Spiking Neuron (MSN), which is implicitly
defined by its electronic circuit.
In the right panel of the figure we
show the experimental neuron equation of state, which is nothing other 
than the activation
function, as noted above. We can observe that near the excitation 
threshold the equation of state is well represented by
the functional form of the activation function of the LIF mathematical 
model (red fitting line in Fig.\ref{fig3}).
At intermediate input currents, the behavior approaches the Ideal IF neuron, 
whose equation of state is $f \propto I$ (see \ref{eq:ideal}), since $f = 1/ISI$

The same methodology allows us to consider more complex 
neuron models, which are also defined by their respective circuit implementation.
For instance, we may consider the case of bursting neurons. From theoretical neuroscience
we know that a requirement to obtain bursting behavior is by adding a second
dynamical variable, besides the potential of the cell body $u(t)$, in Eq.\ref{eq:ideal} and represented by the capacitor in the circuits 
of Fig.\ref{fig2} and \ref{fig3}. 
Thus, to do this we add a second $RC$ pair to our basic MSN 
circuit. A simple option is to add a capacitor $C_L$ in parallel to
the (small) output resistor $R_L$, introducing a new time constant $\tau_L= R_LC_L$.
The resulting circuit is shown in Fig.\ref{fig4}a. In panel (b)
we show the dynamical behavior
that it now produces. We observe four qualitatively different spiking types: 
simple tonic spiking, fast spiking, and two bursting modes. 
These four spiking types are realized in the respective regions of the 
phase diagram of Fig.\ref{fig1} presented before.

\begin{figure}
    \centering
    \includegraphics[width=1\linewidth]{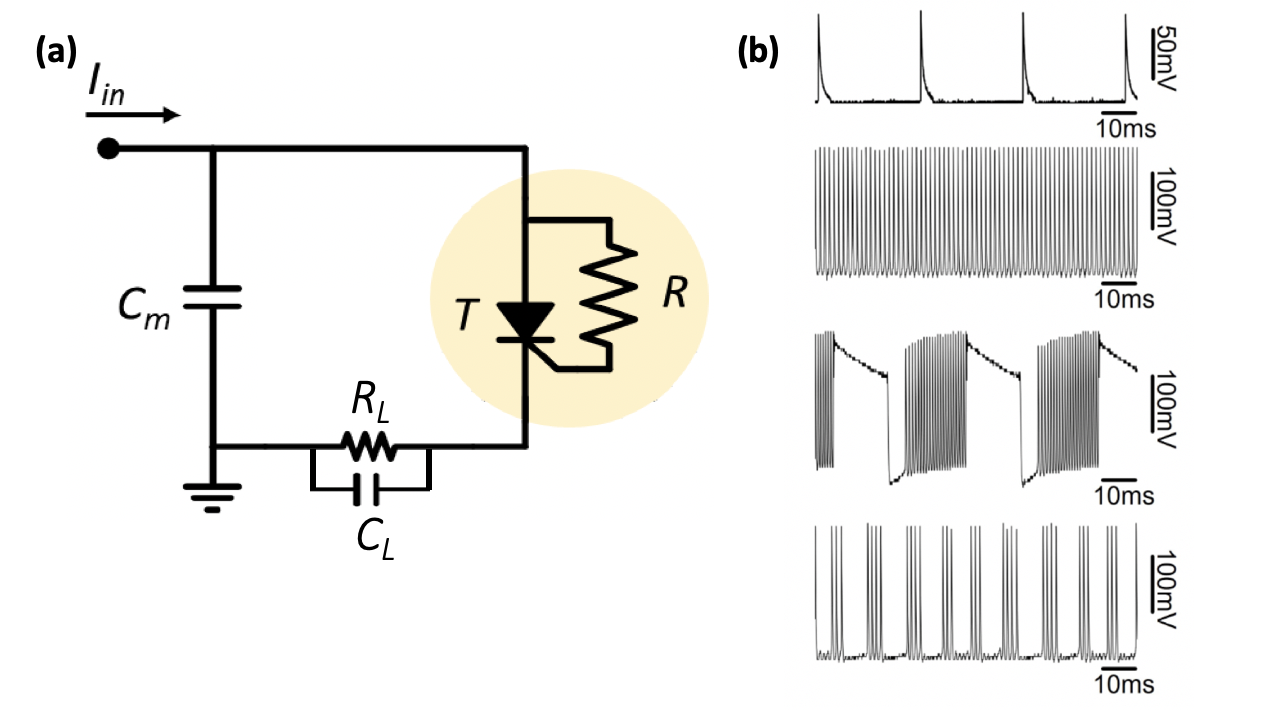}
    \caption{(a) Electric circuit that defines the physical model 
    of a bursting neuron based on the MSN model and adding a second time
    constant $\tau_L = R_LC_L$. 
    (b) The various dynamical behaviors produced by the circuit. From top to bottom:
    tonic spiking (TS), fast spiking (FS) and two bursting types (IB1, IB2),
    which correspond to the four phases of the phase diagram shown in Fig.\ref{fig1}.
    }
\label{fig4}
\end{figure}

\begin{figure}
    \centering
    \includegraphics[width=1\linewidth]{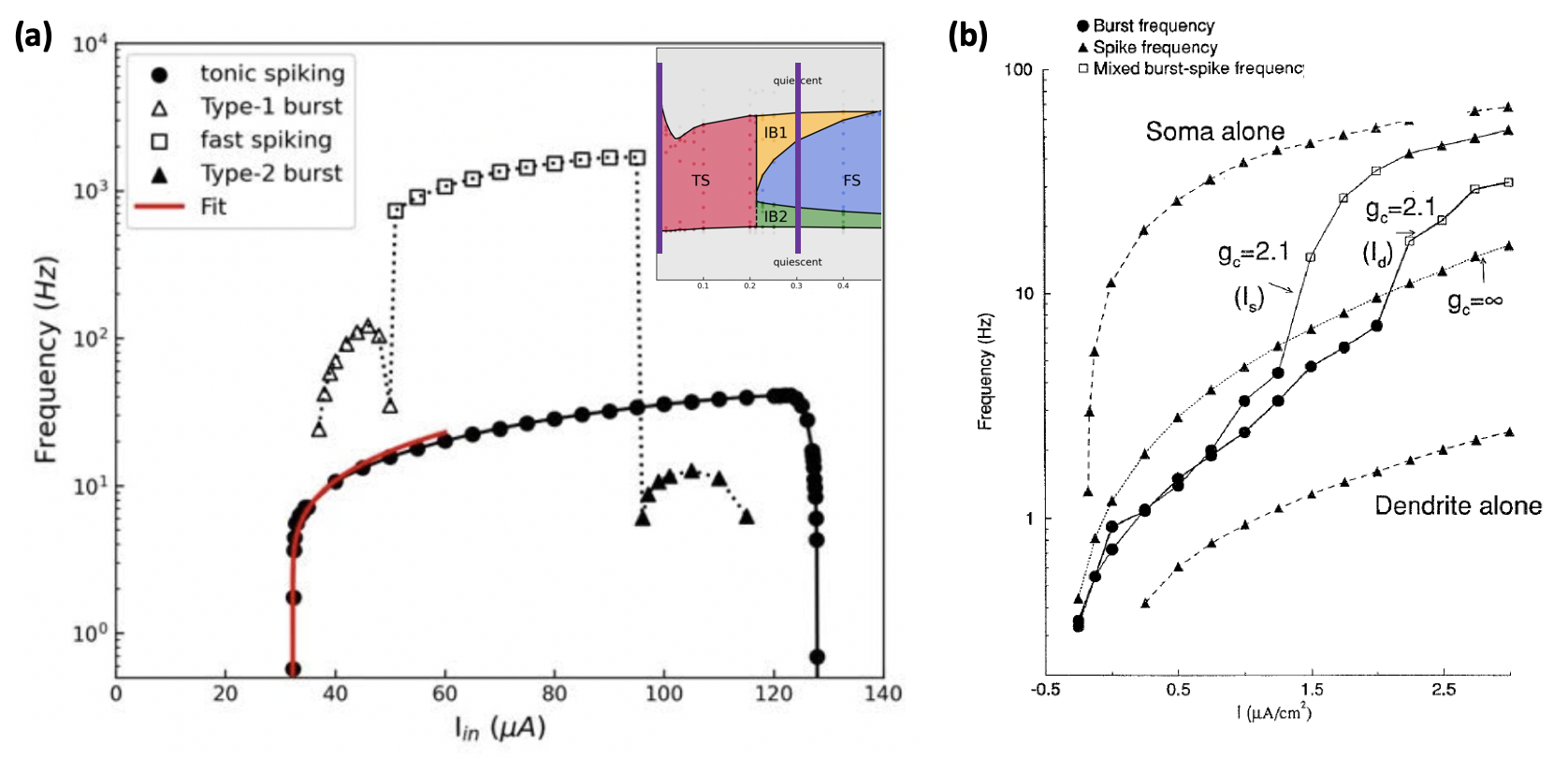}
    \caption{(a) Activation function $f(I)$ in semilog scale. 
    The top curve corresponds to the
    MBN model along the right purple line at $\tau_L$=0.3ms indicated 
    in the phase diagram in the inset. Following the definition of Pinsky and Rinzel,
    the frequency is defined as the inverse of the period between spike trains (bursts).
    The bottom $f(I)$ curve is that of the MSN model discussed before, which corresponds
    to the limit $\tau_L \to 0$ of the bursting neuron model 
    (indicated with the left purple line in the inset). 
    (b) Activation function of the Pinsky-Rinzel model 
    reproduced from \cite{PRmodel}. 
    }
\label{fig5}
\end{figure}

As done before for the basic MSN, 
we may also obtain the equation of state
of the Memristor Bursting Neuron (MBN) model. 
In the present case, we can consider that the time constant $\tau_L$ plays the 
role of a third parameter, similarly as the temperature in the case of 
matter systems.
Hence, in Fig.\ref{fig5} we show the curve $f_{\tau_L}(I)$ measured at a fixed
$\tau_L$, indicated by the vertical purple line that crosses three phases. 
We observe the jumps in the frequency as the current drives the system from
one phase to the other. This is reminiscent of changes in density
when the pressure drives phase transitions at a fixed $T$ in the 
phase diagram of water in Fig.\ref{fig1}.

It is interesting to mention that a biologically realistic
theoretical model of bursting neurons introduced by Pinsky and Rinzel (PR) shows
a qualitatively similar behavior \cite{PRmodel}. 
The PR is an example of a two-compartment model, 
where both the soma and the dendrites are described. We may notice that 
in the MBN model the $R_LC_L$ block (see Fig.\ref{fig4}) can be
considered as a second compartment, which is connected to the output of the
first.
In the right panel of Fig.\ref{fig5}, we reproduce the activation function
(i.e. the neuronal equation of state) of the PR model. 
It is interesting to observe that in the simpler limits of only one 
compartment (soma alone and dendrite alone) the behavior of the PR
is qualitatively the same as that of our basic MSN, which is
also a single compartment model (see
bottom curve of Fig.\ref{fig5}a). 
More importantly, for the relevant case of two compartments (i.e. finite $g_c$) 
we observe a sudden changes in the firing rate as a function of excitatory input current,
also in qualitative agreement with the MBN.
In fact, both PR and MBN traverse the same sequence as the excitatory current is increased: initially quiescent below a critical current, then bursting, and finally
jump in firing rate to the fast spiking mode.
Hence, the phase transitions are abrupt, through a steep
or a discontinuous change in the activation function. 
As we shall discuss in the next section this feature may have interesting
consequences. Moreover, and we shall show that the $f(I)$ anomalies may be 
considered as the counterparts of certain phenomena occurring in 
phase transitions in matter systems. 

\section{Correlation and Response Functions}
Correlation and response functions
are useful concepts in material science, where
they serve to characterize different states of matter.
For instance, a regularity in the arrangement of positions
of atoms is revealed by Bragg peaks in the x-ray spectra, 
which are maxima of the structure factor. 
In real space, the regularity is
revealed by the
pair correlation function  
\begin{equation}
g({x}) = \frac{\int n(x+x') n(x') dx'}{\int n^2(x') dx'}
\end{equation}
where $n(x)$ indicates the particle 
density (such as electrons, atoms, molecules, etc)
at position $x$, and where we consider one dimension 
for simplicity.
In the case of a crystalline order, $g(x)$ shows structure with peaks.
For a simple arrangement of
particles along one dimension with a lattice constant $a$, 
the peaks will be at $a$, $2a$, $3a$, ...
In contrast, for a disordered state, such as gas or liquid,
the $g(x)$ is mostly featureless. The study of $g(x)$ is routinely
done in condensed matter physics for the study of phase 
transitions (see, for example, \cite{liquidNb}).

\begin{figure}
    \centering
    \includegraphics[width=1\linewidth]{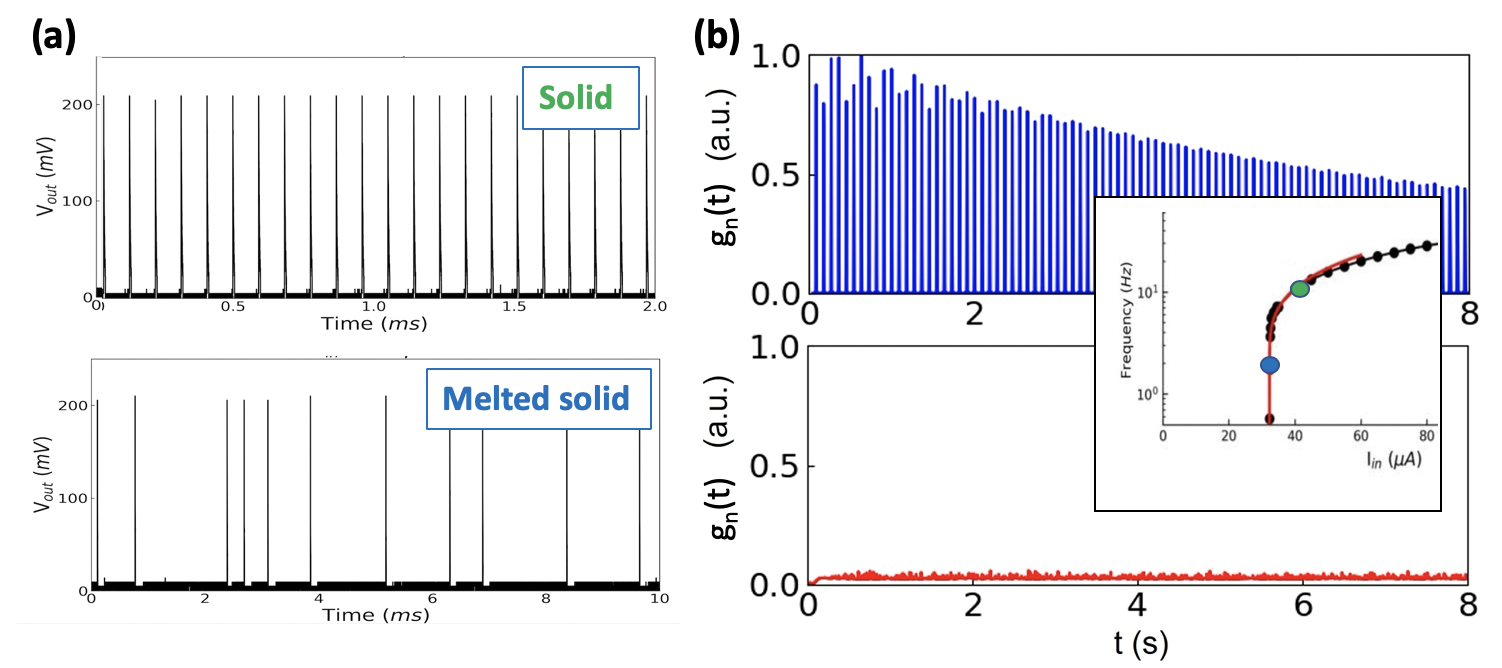}
    \caption{(a) Measured spike traces $s(t)$ of the MSN model
    at two values of the input current: a ``solid'' state measured
    at $I$ = 45.3$\mu$A
    (top) and  a ``melted'' state at $I$ = 36.4$\mu$A. 
    The states are indicated with the green and blue dots in the 
    activation function of the neuron reproduced in the inset.
    (b) The neural correlation function $g_n(t)$ computed for the respective 
    traces shown in the left side panels (only a small portion
    of the measured traces is shown). 
    }
\label{fig6}
\end{figure}

In our analogy, spiking systems are thought of as temporal-matter states,
so it is natural to explore the behavior of the correlation-function analogue to $g(x)$.
Since the position of particles correspond to
position of spikes, by analogy we can
define the neural correlation function $g_n(t)$ as
\begin{equation}
g_n(t) = \frac{\int s(t+t') s(t') dt'}{\int s^2(t') dt'}
\end{equation}
where $s(t)$ indicates a given spike trace. 
The function $g_n(t)$ can
characterize different spiking states of a neuron.
In Fig.\ref{fig6} we provide a concrete example, which is realized in 
the basic MSN model described before (see Fig.\ref{fig3}).

We can observe the two qualitatively different tonic spiking behaviors in the two left panels
of the figure. They correspond to two constant current inputs (36.4$\mu$A and 45.3$\mu$A).
In the first case, at higher input current (top panel), the spiking is perfectly regular.
In contrast for a smaller current close to the threshold, 
the trace changes dramatically. 
The interval between spikes become very irregular. 
By analogy between spike and particles, we can think of the
first case as that of a solid and the second as of the melting of the solid state. 
This qualitative description can be made more precise by the correlation
function $g_n(t)$, shown on the right side panels of Fig.\ref{fig6}.

The top panel shows a succession 
    of delta functions at equally spaced at times, multiples of the constant 
    inter-spike interval, $t_k$=$k$ISI.
    This indicates the long range order in time. It tells that given the
    presence of a spike at time $t=0$ we have a high probability (~ 1)
    of finding another spike event at times $t_k$ ($k$=1, 2, 3, ...).
    In contrast, the $g_n(t)$ shown in the bottom panel is featureless, 
    showing a small approximate constant value. This  
    indicates total lack of order as the presence of a spike a time
    $t=0$ does not permit to predict the presence of ulterior spikes
    events. The emission of spikes is random as the positions of particles 
    in a liquid or a gas.

The peaks of the $g_n(t)$ are very narrow, delta-function-like, 
because the spikes are very narrow with respect to the duration of the ISI. 
In a solid, the atoms have a size that is smaller but of the
same order as the lattice spacing, so instead of narrow deltas 
one observes broad peaks in the $g(x)$ \cite{liquidNb}.

One may understand quite intuitively the physical origin of this 
dramatic changes in the time-structure. 
The key point is to realize that
the ``melted'' state occurs in a regime where the activation 
function $f(I)$ is very steep, at the onset of neuron excitability,
i.e. near the threshold.
Therefore, small variations of the input current 
will reflect on significant variations in the ISI.
This observation motivates the following important insight. 

This enhanced sensitivity to current fluctuations
is due to the large slope $df/dI$ in the activation function.
Then, what is this feature related to, if one follows the analogy
back to the matter systems?
We recall that ISI plays the role of the lattice spacing, 
or specific volume (see Eqs.\ref{eq:ideal} and
\ref{eq:equiv}), then $f$=1/ISI corresponds the particle 
density $n(x) = 1/v(x)$. On the other hand, since the 
input current $I$ is like the pressure, 
then it follows that the slope $df/dI$ corresponds to 
$dn/dP$. This last quantity is
closely related to the compressibility of matter systems
\begin{equation}
    \beta=\frac{1}{n} \ ( \frac{dn}{dP} )
\end{equation}
which is the inverse of the bulk modulus.
We can therefore follow the analogy and 
introduce the concept of ``neuro-compressibility'',
\begin{equation}
    \beta_n = \frac{1}{f} \ ( \frac{df}{dI} )
    \label{neural_comp}
\end{equation}

It is important to mention that this quantity may be measured using 
experimental methods such as dynamic clamp, 
where a controlled synaptic current can be
injected into a neuron while its activity is monitored \cite{dynamic_clamp}.
Moreover, this definition may turn out to have important consequences, as 
we discuss next. 

Anomalies in the compressibility 
of materials are precursor signatures of structural phase transitions. 
A sudden increase in the
compressibility of a solid indicates the ``softening'' of a 
vibrational mode (a phonon mode), which 
leads to a change in the structure, or possibly a phase change.
Then, the question is, what would the analogue phenomenon be for a neuronal system?
For a single neuron, the enhancement of $\beta_n$ would indicate the proximity to
a qualitative change in the spiking mode of a neuron,
i.e. a ``bifurcation'' in its dynamics. This can in fact be seen in the panels
of Fig.\ref{fig5}. 
There, we observe that all the changes in the spiking modes for both, the 
MSN and for the MBN models, occur at current values where there are enhancements 
in $df/dI$ or jumps. 
Most notably, this is not only a feature of our artificial neuron circuits, but also
can be clearly seen in the biologically realistic Pinsky-Rinzel model activation functions
that we reproduced in Fig.\ref{fig5} \cite{PRmodel}.
The enhancements seen in the PR data occur at the onset of change from
quiescent to spiking and also in the change from burst to tonic spiking 
(black circles and black triangles), in very good qualitative agreement 
to our electronic neuron model,

We may then speculate on an important implication of our observations. 
It would be interesting to explore if neuro-compressibility anomalies are also 
found across the boundaries of qualitatively different  states in 
{\it neuron networks}. 
If that is the case, an intriguing and exciting possibility would be to investigate 
if anomalies in $\beta_n$ are also detected (by small current stimulation) in 
animal models of epilepsy and Parkinson's disease. 
If this were the case, then one may envision a pathway to a novel diagnose 
tool for early detection or a risk predictor of mental diseases associated
to abnormal spike patterns in humans. In even further speculation, 
one may also search for anomalies at the onset of regaining, or loosing, 
consciousness, which is another challenging frontier of research 
\cite{regaining_cons}. 

\section{Bursting spikes as an ananlogue of formation of clusters defects}

Here we describe another interesting connection between common a phenomena 
in spiking neurons and in material science.
We shall show that missing spikes in a trace of a fast spiking state, 
can be thought of as the analogue of missing atoms, i.e, defects, in a crystal 
structure. 

\begin{figure}
    \centering
    \includegraphics[width=1\linewidth]{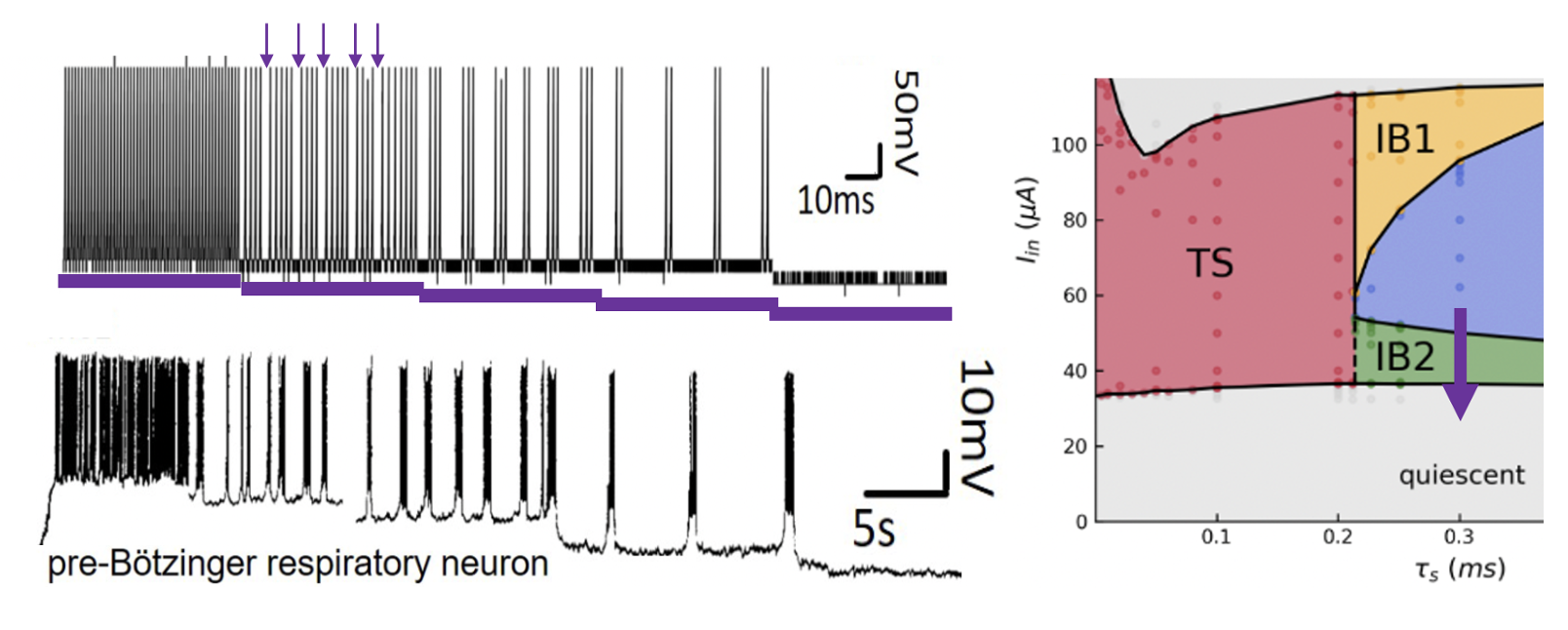}
    \caption{Top left: Measured spike traces of the MBN
    as a function of decreasing current in discrete steps (purple). 
    The thin purple arrows indicate the missing spikes. 
    Bottom left: Experimental trace of pre-B\"otzinger bursting neurons 
    from rats. The data is obtained by changing excitatory input current in discrete 
    steps. Adapted from \cite{pre-botz}. The circuit parameters may be
    easily adjusted to fit the experimental data \cite{MBN}.
    Right: Phase diagram of the MBN 
    where the purple arrow indicates the evolution from the fast spiking (blue)
    to the bursting type 1 phase (green) by decreasing current at constant $\tau_L$.
    }
\label{fig:defects}
\end{figure}

From Fig.\ref{fig:defects} we observe that the proliferation of missing 
spikes in a fast spiking state is a route to generate bursting behavior. 
This is illustrated in the sequence of traces shown in the figure, 
which were obtained for a step-wise decreasing input current to the MBN.  
The thick purple arrow in the vertical path followed in phase diagram
(from blue to green to grey).
In the top trace we indicate with small purple arrows the missing arrows, 
showing that one may understand the onset of bursting as the result of 
skipping spike events, which are initially few (i.e. dilute). 
As the current intensity is further reduced, 
the  missing spikes become more numerous (i.e dense) and occurring in
{\it clusters} of inactivity, which give rise to the stuttering mode 
bursts \cite{stuttering}. 
In our analogy, we think of spikes as of atoms in a lattice, therefore, the 
initial continuous fast spiking state is like a perfect crystal. 
The missing spikes then play the analogue role as vacancy defects, i.e.
lacking atoms. Moreover, the missing spikes are
the result of decreasing current, which in the analogy represents pressure. 
It is then interesting to observe that in thin-film deposition, which is
a topic in material science, the partial pressure of oxygen $P$(O$_2$)
is a relevant parameter for the quality of the growth of crystalline oxides. 
Moreover, it is well known that reducing the $P$(O$_2$) 
induces the creation of oxygen vacancy defects in the crystal
structure \cite{tiO2,ceo2}, which often cluster together forming dislocations \cite{sto}, 
This is in full qualitative analogy to the spiking traces in the stuttering
bursting mode shown in Fig.\ref{fig:defects}.

We would like to emphasize that the path of phase transformations, i.e.
the evolution from fast spiking, to bursting, to quiescent does not seem to
be just a peculiarity of our MBN circuit model. In the lower panel of the
Fig.\ref{fig:defects} we illustrate the striking resemblance of the traces
of the MBN with those measured in bursting neuron of rats \cite{pre-botz}.
Quite remarkably, the experimental traces were obtained by solely changing the
intensity of the excitatory DC current. 

\section{Neural Networks}

We now consider one important final aspect in our analogy that may eventually 
bring new light to the issue of how to think about inter-neuron coupling. 
So far we have considered essentially individual neurons, but we may ask
what would it be to extend the analogy to multi-neuron systems, i.e. to 
neural networks.

\begin{figure}
%    \centering
    \includegraphics[width=\linewidth]{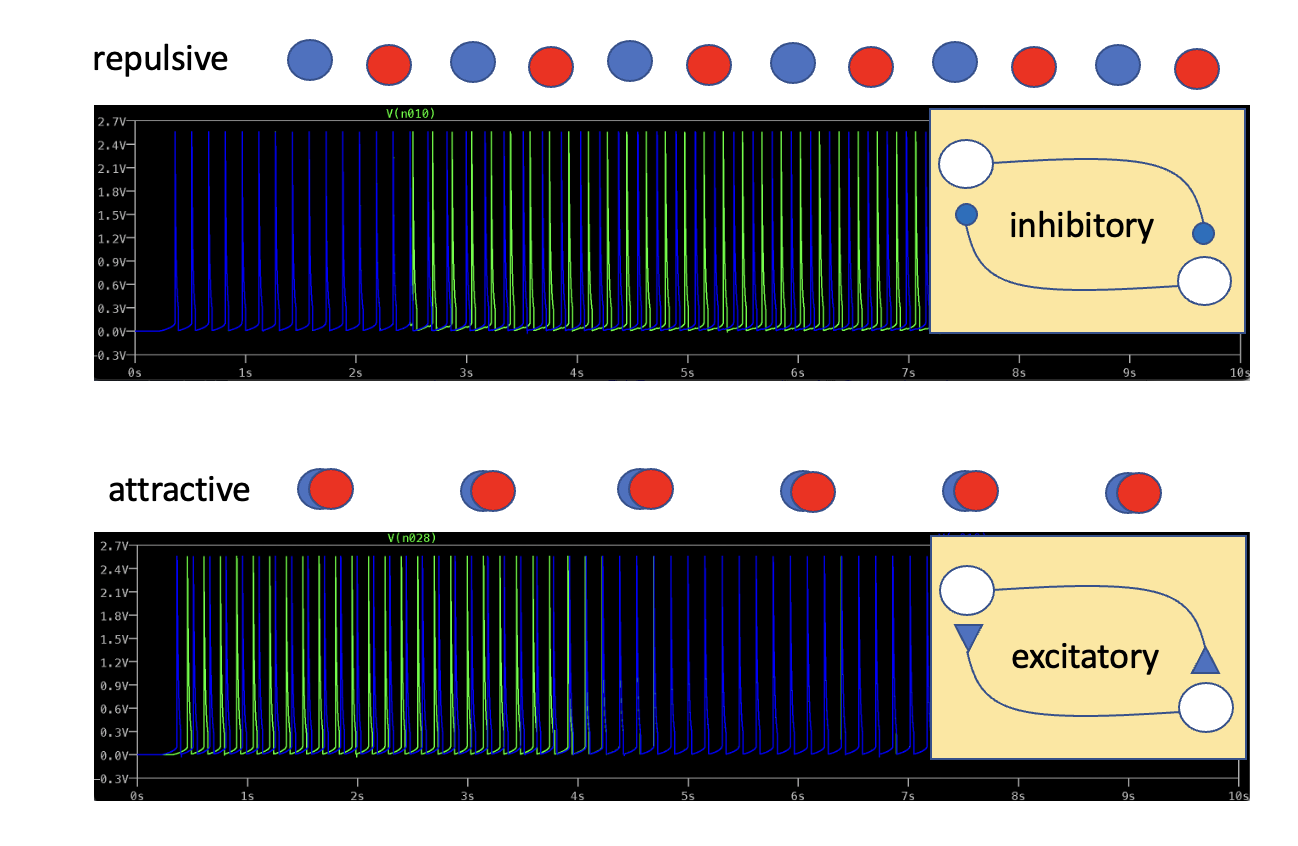}
    \caption{Top: two 1D space lattices (red and blue) with repulsive interaction and two neurons
    with inhibitory interactions. The neurons are prepared to start spiking synchronously and spontaneously
    adopt an alternating spiking pattern. Bottom: two 1D lattices with attractive interactions and two
    neuron network with excitatory interaction. They are prepared to alternate but spontaneously begin
    to spike in unison. Spike traces were produced by realistic simulations
    with LTSpice.}
    \label{fig7}
\end{figure}

As a first glimpse into this question, we shall consider the simplest network case,
namely, just two neurons that are mutually excitatory or inhibitory. 
We focus first in a system of two identical neurons, each excited with equal 
input currents.
The currents are above threshold, so the neurons are active and theirs spikes are
transformed via conductances into mutually injected synaptic currents that are positive
for the excitatory case and negative for the inhibitory one. 
This is schematically shown in Fig.\ref{fig7}.

We shall see that our analogy takes an interesting twist, as the dynamical
states of the two-neuron system can be considered as an analogue of 
a complex crystal, i.e. a crystal with two atoms in the unit cell. 
Moreover, the coupling between neurons is mediated by synaptic currents, 
which can be excitatory
or inhibitory. Since in our analogy current plays the role of pressure, 
then the synaptic currents should also play such a role. 
More precisely, an excitatory synaptic current should correspond to
a repulsive inter-particle interactions (positive pressure), and an inhibitory 
synaptic current should be an attractive interaction (negative pressure).  
We shall consider the two cases, which we study using realistic electronic 
circuit simulations (LTspice).

We first consider two spiking neurons with mutual inhibition. When one neuron spikes it inhibits 
the other, and vise-versa, so they try to avoid spiking at unison since we are synaptic current are
instantaneous. After a transient period, they expectedly find a stable dynamical state where they 
alternate to emit spikes 
as shown in the top panel of Fig.\ref{fig7}.
Using our analogy, this spiking 
pattern corresponds two a perfect molecular crystal where the unit cell has an A-B atom pair 
(or a basis).  

In the bottom panel of the figure we consider the second case, that of mutually 
excitatory neurons. Again after a transient time, the system adopts a periodic spiking
pattern. However, in contrast to the previous case, now both neurons
fire at unison. This is also intuitive, the excitatory synaptic current of
emitted by a neuron that spiked promotes the spiking of the other one, and vise versa. 
So, naturally, the both spike at the same time, which is nothing else that
excitatory synapses promote synchrony in neural networks \cite{synch}. 
Following our analogy, spiking at unison corresponds
to a ``dimerization'' of the lattice. Namely, that the distance between the 
A-B pair atoms is reduced as due to an attractive interaction between the A and B
atomic species. 

These two cases are consistent with our analogy, where current is interpreted
as pressure. Indeed, the volume of the unit cell in the inhibitory case is large, 
as expected for positive effective pressure between A and B, while in the 
excitatory case the volume is fully collapsed to zero, as expected for
a negative pressure within the unit cell.

It is an interesting perspective for future work to consider increasingly complex
networks of several neurons (motifs). The periodic states that emerge constitute
spiking {\it sequences}, which are of great relevance for automatic motor behavior. 
By virtue of the analogy that we introduced in the present work, those periodic sequences
should correspond to a variety of molecular crystals. 
It would be exciting to explore if new intuition for Neuroscience could be brought
from from those traditional areas of Condensed Matter Physics and Chemistry \cite{molecX-tal}.

\section{Conclusion}

In this work we introduced the idea that the dynamical states of neural networks
may be though of as realizations of ``time matter'' states.

We started from the notion that a trace of spiking events of a neuron can be analogue to 
a snapshot of particles or atoms arranged in space. We then went on to explore and
show that the analogy may be pushed far beyond that literary statement, and may
provide new intuition in the challenging problem
of understanding and designing spiking neural networks.

We identified analogue roles of basic quantities of Physics with those of Neuroscience, 
such as pressure and volume with input currents and inter-spike intervals. 
We then logically build on this assumption to show connections between correlation 
and response functions in both fields.
Perhaps most significant was the finding that a {\it neuro-compressibility} can be 
defined, with possible far reaching consequences, including medical ones, that may be 
experimentally tested. 

An exciting new road of discovery may open ahead.

%\maketitle

%\tableofcontents

%\section{\label{sec:level1}First-level heading:\protect\\ The line
%break was forced \lowercase{via} \textbackslash\textbackslash}

\section{Acknowledgments}
We acknowledge support from the French 
ANR ``MoMA'' project ANR-19-CE30-0020. 

\bibliography{general_bibl}

\end{document}